\documentclass[twocolumn]{aastex63}

\usepackage{amssymb}
\usepackage{multirow}

\usepackage{xcolor}
\newcommand{\report}[1]{\textcolor{black}{#1}}

\usepackage{changepage}

\received{MM DD, 2021}
\revised{MM DD, 2021}
\accepted{MM DD, 2021}
\submitjournal{ApJS}

\shorttitle{Classification of OGLE eclipsing binary stars}
\shortauthors{B\'odi \& Hajdu}

\begin{document}

\title{Classification of OGLE eclipsing binary stars based on their morphology type with Locally Linear Embedding}

\author[0000-0002-8585-4544]{A. B\'odi}
\email{bodi.attila@csfk.org}
\affiliation{Konkoly Observatory, Research Centre for Astronomy and Earth Sciences, E\"otv\"os Lor\'and Research Network (ELKH)\\
H-1121 Budapest, Konkoly Thege Mikl\'os \'ut 15-17, Hungary\\}
\affiliation{MTA CSFK Lend\"ulet Near-Field Cosmology Research Group\\}

\author[0000-0001-8060-2367]{T. Hajdu}
\affiliation{Konkoly Observatory, Research Centre for Astronomy and Earth Sciences, E\"otv\"os Lor\'and Research Network (ELKH)\\
H-1121 Budapest, Konkoly Thege Mikl\'os \'ut 15-17, Hungary\\}
\affiliation{MTA CSFK Lend\"ulet Near-Field Cosmology Research Group\\}
\affiliation{E\"otv\"os Lor\'and University, P\'azm\'any P\'eter s\'et\'any 1/A, Budapest, Hungary}




\begin{abstract}
The Optical Gravitational Lensing Experiment (OGLE) continuously monitors hundreds of thousands of eclipsing binaries in the field of galactic bulge and the Magellanic Clouds. These objects have been classified into main morphological sub-classes, such as contact, non-contact, ellipsoidal and cataclysmic variables, both by matching the light curves with predefined templates and visual inspections. Here we present the result of a machine learned automatic classification based on the morphology of light curves inspired by the classification of eclipsing binaries observed by the original Kepler mission. We similarly use a dimensionality reduction technique, the locally linear embedding to map the high dimension of the data set into a low dimensional embedding parameter space, while keeping the local geometry and the similarities of the neighbouring data points. After three consecutive steps, we assign one parameter to each binary star, which well scales with the ``detachness'', i.e. the sum of the relative radii of the components. This value is in good agreement with the morphology types listed in the OGLE catalog and, along with the orbital periods, can be used to filter any morphological sub-types based on the similarity of light curves. Our open-source pipeline can be applied in a fully automatic way to any other large data set to classify binary stars.

\end{abstract}



\section{Introduction}
\label{sec:intro}

Eclipsing binary stars are import astrophysical testbeds as their fundamental parameters such as masses, radii, temperatures, absolute luminosities, and rotation can be measured directly \citep{Andersen91,WD71,Wilson12,Phoebe,Phoebe20}. Stellar and galactic evolutionary models can be tested using these objects, because the components have the same age and chemical composition \citep{EBevol}. Binary stars are also useful for studying the theory of stellar interaction between the components, mass exchange and mass loss mechanisms, limb darkening, magnetic activity and measuring distances. Thanks to their stable orbits, they operate like precise clocks on the sky, which permits the discovery of dynamical evolutionary effects \citep{EBdin1,EBdin2,EBdin3,Naoz16,Moe18,Hamers21} or finding new companions orbiting around the binary's center of mass \citep{Borkovits2015,Borkovits2016, ZacheLMC,ZacheSMC,HajduTriples,Mitnyan20}.

The Optical Gravitational Lensing Experiment (OGLE; \citealt{Udalski2015}), which observes the Milky Way, Small- and Large Magellanic Clouds, discovered hundreds of thousands of eclipsing binary stars. The OGLE collection of eclipsing and ellipsoidal binary stars currently contains more than 48~000 systems in the Magellanic Clouds \citep{OGLEMC} and over 450~000 binaries in the field of galactic bulge \citep{OGLEBLG}. The observations are collected in two optical photometric bands, Johnson V and \report{Cousins} I bands, of which the latter is the default with about \report{15} times more epochs than the former. The obtained light curves mostly have high signal-to-noise ratios and their types are confirmed by experts, which makes the sample very reliable. The catalog consist of all kind of binary star types from detached, semi-detached and contact systems to more exotic objects such as RS CVn stars, cataclysmic variables, HW Vir binaries, double periodic variables and even some planetary transits. These samples are characterized by exceptionally high levels of completeness and purity, therefore they serve as a framework for various astrophysical applications. The fourth phase of the survey (OGLE-IV) has started in 2010, but due to the global pandemic of coronavirus disease 2019 (SARS-CoV-2), the regular observations has stopped \report{on 18 March 2020}.

The OGLE eclipsing binaries have been selected and classified in a two step approach \citep{OGLEeclClass}. First of all, an extensive period search has been conducted using Lomb-Scargle periodogram \citep{Scargle82} and the Box-Least Squares (BLS) period-search algorithm \citep{BLSalgo}. While the former is sensitive to sinusoidal light curve shapes, the latter capable of finding box-shaped periodic signals, which is common in detached eclipsing binaries. Due to the large number of stars, a Random Forest (RF; \citealt{RF}) algorithm has been trained on the BLS statistics using the best-sampled OGLE-IV fields as a training set. The RF classifier yielded the first classification labels. The second step was to phase fold and normalize each light curve into the range of 0-1 and cross-correlate them with a collection of template curves. Finally, all light curves have been visually inspected at least once and obvious non-eclipsing stars were removed from the sample.

Several attempts have already been made to automatically classify light curves in the OGLE database \citep{Wyrzykowski04,Sarro09,Richards11,Pawlak13,Aguirre19,Szklenar20,Becker20}. These authors use supervised machine learning algorithms to separate binary stars from other variable star classes or to assign one of the four main morphological categories to each binary star. Although \citet{OGLEunsupervised} tested their unsupervised (i.e. non-labelled) algorithm on OGLE-III data set, they only considered binaries as one of the classes to be separated from others.
The first unsupervised classification based on similarity of light curve of eclipsing binary stars was performed by \citet{KeplerLLE}. They used locally linear embedding (LLE) algorithm \citep{LLEalgo} to assign continuously measured values to stars observed by the original Kepler mission \citep{KeplerMission}. Later \citet{K2LLE} used the same method to separate detached, semidetached and
contact eclipsing binaries observed by the K2 mission in its campaigns 0-4. The Gaia eclipsing binary and multiple systems have also been classified based on their light curve morphology using a supervised classification tool, linear discriminant analysis \citep{LinearAnalysis}, which provides a similar morphology parameter as the LLE for Kepler targets \citep{GaiaEBclass1,GaiaEBclass2}. Although these authors aimed to use machine learning techniques to classify binary stars into a small number of distinct groups, these algorithms have a great advantage, besides the speed of the process, which is almost not comparable to the human inspected way, they do not depend on human impressions, which may lead to unreliable results. Moreover, similar eclipsing binary light curves can be more efficiently selected for well-aimed studies if a continuously measured number is assigned to each of them.

The goal of this paper is to use the LLE algorithm for the purpose of classifying the OGLE eclipsing binary light curves based on their morphology. To do so, we first refine the times of minimum brightness and periods listed in the OGLE catalog, then fit a polynomial chain to the phased curves which are re-sampled and re-scaled before the classification.

This paper is structured as follows: In Section\,\ref{sec:methods} we describe the data preparation steps, and the used algorithms, in Section\,\ref{sec:results} we present the classification results and in Section\,\ref{sec:summary} we draw a brief conclusion.


\section{Data and methods}
\label{sec:methods}

\subsection{Data preparation}
\label{sec:preparation}

We downloaded all OGLE-IV I-band eclipsing binary light curves obtained in the field of galactic bulge and the Magellanic Clouds from the OGLE database\footnote{http://www.astrouw.edu.pl/ogle/}. However, not all listed targets have available measurements in the fourth phase. In this cases we complemented the data set with OGLE-III light curves. Out of the cataloged 450\,598, 40\,204 and 8\,401 objects in the Galactic Bulge, Large- and Small Magellanic Clouds, respectively, 892, 41 and 2 stars have no measurements or the phase coverage is so poor that we were not able to fit the light curve, thus these objects were excluded from our analysis.

The OGLE catalog also lists some basic properties such as the coordinates, out-of-eclipse magnitudes of the systems and, most importantly, the epochs (T$_0$), periods (P), and depths of both minima. The T$_0$ and P values can be used to phase fold light curves centering the primary minimum at zero phase. However, we found that in several cases the T$_0$ is misaligned, and fine-tuning the period may decrease the scatter of the phase curves. Therefore, we adjusted the T$_0$ values by shifting the primary minima of the fitted polynomial chain models (see later) to zero phase. To refine the periods we used the \report{Phase Dispersion Method (PDM; \citealt{Stellingwerf78})}, which calculates the scatter in each bin of phase folded and binned light curves and minimize this value with respect to the scatter of the full phase curve by adjusting the period. We assumed that the OGLE period is close to the true value, which made possible to tighten the period search range to $\pm0.01$ days around the initial value. As the precision of the period is dependent on the number of period samplings, the finer the period grid the preciser the result, we used the PDM algorithm implemented in the GPU-based Python code \texttt{cuvarbase}\footnote{https://johnh2o2.github.io/cuvarbase/}, which utilize the speed of graphics processing units. This way the time required to calculate a spectrum on a fine grid reduced significantly. For each star, we defined the grid in frequency range with 100,000 equidistant steps.

During the PDM analysis, another important parameter, which has to be set by the user, is the number of phase bins. As choosing an inappropriate value may smear out the eclipses, we scaled this parameter taking into account the Kepler law in the following way.

\report{In the edge-on system,} the semi-major axis of a binary star can be written as
\begin{equation}
a = \sqrt[3]{\frac{GMP^2}{2\pi^2}},
\end{equation}
where $G$ is the gravitational constant, $M$ is the total mass, $P$ is the orbital period. Assuming a circular orbit, the angle between the first and fourth contacts
\begin{equation}
\alpha = \arctan{\bigg(\frac{R}{a}\bigg)}^2,
\end{equation}
of which the eclipse duration
\begin{equation}
\Delta t = \frac{2 \alpha P}{2\pi}.
\end{equation}
The ``ideal" number of bins to sample the whole eclipse with a high enough resolution can be written as
\begin{equation}
n_{\mathrm{bins}} = \frac{10}{\Delta t}P.
\end{equation}

To estimate the bin numbers, we assumed that both components have one solar mass and radius. As the resulted value can be too small or too large for very short or very long periods, respectively, we prohibited the number of bins to be smaller than 50 or larger than 2000.

It is common that outliers appear in the light curves, and these points may have significant influence on the model fittings. To identify and remove these points, we used the DBSCAN (Density Based Spatial Clustering of Applications with Noise; \citealt{DBSCAN}) algorithm implemented in \texttt{scikit-learn}. First, we phase folded the light curves to increase the density of measurements, then scaled the \report{peak-to-peak} brightness into the range of 0--0.3 or 0--0.5 for stars with orbital period less than or longer than 5 days, respectively. This vertical scaling was used to transform the area defined by the maximum distance between adjacent points -- used for the clustering -- into an ellipsoid, as we assumed that the points that are part of the eclipses are ordered vertically. Finally, we clustered the points based on the average distance between consecutive points into different classes containing at least two samples. The single points that were labeled as noise were removed.

Finally, we fitted the cleaned phase curves with a chain of four second/fourth order polynomials (\textit{polyfit}, \citealt{polyfit}) connected at a given set of knots. For each star we fitted several models and selected the best one (see Sect.~\ref{sec:polyfit}). As the unsupervised classification would be lead by the depth of minima, instead of its shape, if we were using the raw fits, we re-scaled each model into the range of 0-1 vertically. We use these \textit{polyfit} representations as the input of the classification step described in Sect.~\ref{sec:LLE}.

\subsection{Choosing the best polynomial chain representation}
\label{sec:polyfit}

The shape of light curves varies from the sinusoidal-like W~UMas to the Algol's that show flat maxima with sudden sharp dips, the eclipses. Thus, fitting each data set requires different polynomial orders with very different fix points between the chain links. As the result of \textit{polyfit} model fitting highly dependent on the chosen initial phase location of knots and the fixed polynomial order, we fitted a variety of models. We tested second and fourth order polynomials, where the former is usually sufficient to fit sinusoidal shaped light curves, while the latter is needed to fit sharp eclipses. The initial location of knots were chosen in two ways, firstly, an automatic search was carried out to locate points below the mean of the light curve by 0.5 standard deviation, secondly, we predefined typical values both for W~UMas and Algols. To select the best representation, we used Bayesian Information Criterion (BIC). Assuming that we have normally distributed errors, the BIC can be derived from $\chi^2$ statistics
\begin{equation}
    \chi^2 = \sum_{i=0}^n \frac{(m - \hat{m})^2}{\sigma^2},
\end{equation}
of which
\begin{equation}
    BIC = \chi^2 + k \ln(n),
\end{equation}
where $m$ is the observed magnitudes, $\hat{m}$ is the \textit{polyfit} model, $n$ is the number of observations and $k$ is number of model parameters.

During the polynomial chain regression the location of the knots are updated iteratively for a given number of steps. We limited these steps to a maximum number of 4000, but to further decrease the running time, we monitored the variation of the $\chi2$ value. After 200 steps the mean and the median absolute deviation (MAD) are calculated and after this step if a minimum was found with a $\chi2$ that is lower than the mean minus five times the MAD, the iteration terminated. Moreover, we parallelized the independent fittings and after all processes have had finished, the model with the lowest BIC were selected. We also implemented several additional \report{criteria} to always select that \textit{polyfit} model that most resembles to a simple, but physically expected binary model, e.g. in case of ellipsoidal binaries that also show pulsation, which objects can be recognized by the poor goodness of fits due to the large internal scatter of the phase curves, i.e. very high $\chi^2$ values, we only used the \textit{polyfits} with second order polynomials to not follow the irregular variations.

 \begin{figure*}
   \centering
   \includegraphics[width=\textwidth]{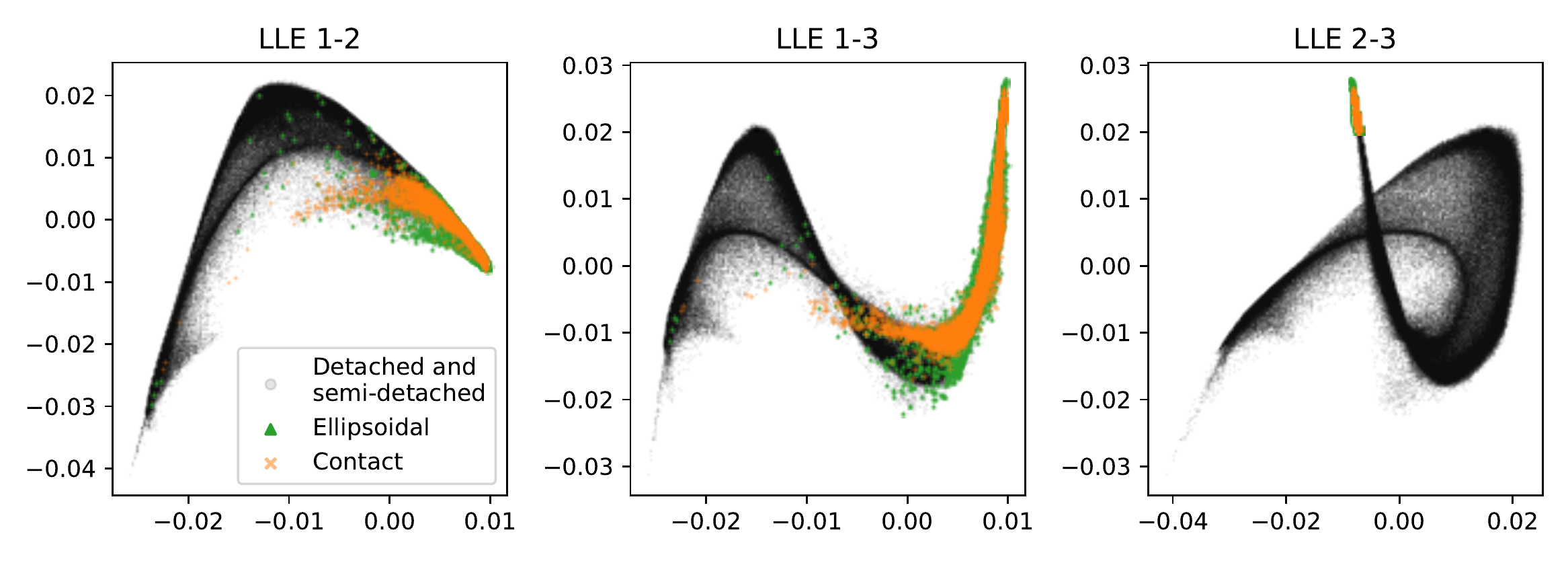}
   \includegraphics[width=\textwidth]{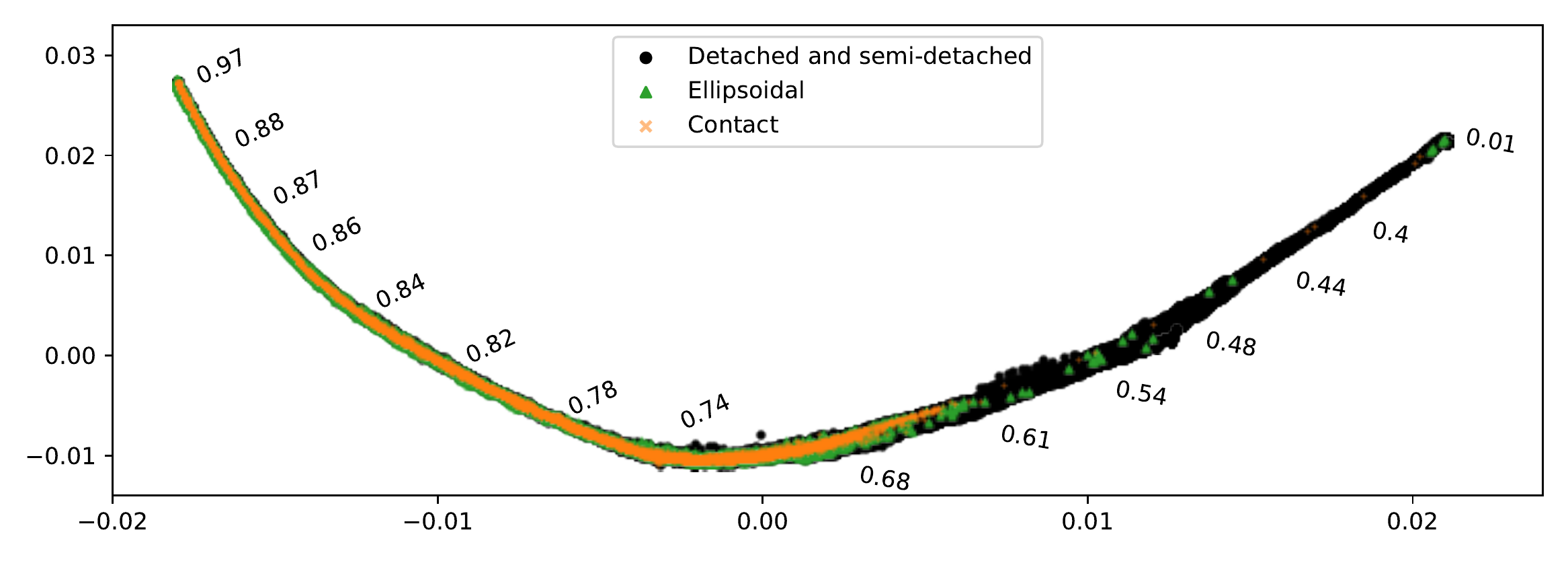}
      \caption{Low dimensional representation of the 1000-bin-long \texttt{polyfit} light curve models. \textit{Upper row}: Three projections of the 3 dimensional Local Linear Embedding manifold of the whole OGLE-IV eclipsing binary data set in the field of galactic bulge. The LLE was performed with nearest neighbours $k=20$ and regularization parameter $r=1.5\cdot10^{-2}$. Different symbols depict the location of stars belonging to the main morphological types listed in the OGLE catalog. \textit{Bottom}: The 2D LLE manifold of the points in the upper panel. The numbers show the approximate position of the morphology parameter $c$ assigned to each binary star with different ``detachness".}
         \label{fig:3DLLE}
   \end{figure*}

\subsection{Local Linear Embedding}
\label{sec:LLE}

The Local Linear Embedding \report{(LLE; \citealt{Chen2011,LI2019336})} method, which was introduced by \citet{LLEalgo}, is an unsupervised machine learning algorithm designed to perform nonlinear dimension reduction. LLE maps a high dimensional data into low dimensional parameter space preserving the local similarities of the original feature space by keeping all the neighbors close to each other.  If the dataset consists of N real-valued, D-dimension vectors $X_i$, which are sampled from an underlying smooth manifold, LLE can reconstruct each point from its neighbouring points. First, for each data point K nearest neighbours are selected to represent each point by the linear combination of its neighbors, then the following cost function is minimized:

\begin{equation} \label{eqLL1}
    \varepsilon(W) = \sum_{i} |\vec{X}_i - \sum_j W_{ij}\vec{X}_{j}|^2,
\end{equation}
where $W_{ij}$ weights are the contribution of $j$th data points to the $i$th reconstruction. The minimization of Eq. \ref{eqLL1} is performed subject to the constrain:
\begin{equation} \label{Wsum1}
    \sum_j W_j = 1,
\end{equation}
which can exploited to rewrite the $i$th component of Eq.~\ref{eqLL1} to the following form:
\begin{equation}
    \varepsilon^i(W) = \sum_{j} |W_{ij} (\vec{X}_i - \vec{X}_j)|^2 =
    \sum_{j=1}^k \sum_{l=1}^k W_{j} W_{l} C_{jl},
\end{equation}
where
\begin{equation}
    C_{jl} = (\vec{X}_i - \vec{X}_j)^T(\vec{X}_i - \vec{X}_l)
\end{equation}
is the correlation matrix of neighborhood points. Now, the optimal weights are
\begin{equation}
    W_{j} = \frac{\sum_l C_{jl}^{-1}}{\sum_m \sum_n C_{mn}^{-1} }.
\end{equation}
In practice, instead of inverting matrix $\mathbf{C}$, the following equation is solved:
\begin{equation}
    \sum_{j} C_{jl} W_l = 1.
\end{equation}
The yielded weight matrix can be re-scaled to match the condition in Eq. \ref{Wsum1}. As the correlation matrix can be singular, which means that the $W_i$ weights would be ill-conditioned, a small multiple of the identity matrix is added to matrix $\mathbf{C}$
\begin{equation}
    \mathbf{C} \rightarrow \mathbf{C} + r\mathbf{I},
\end{equation}
where $r$ is the regularization parameter.

After the $W_{ij}$ weights are calculated, they are used to map the high dimensional observations $\vec{X}_i$ to a low dimensional vector $\vec{Y}_i$. Finally, the lower, $d$-dimensional coordinates $\vec{Y}_i$ are chosen to minimize the embedding cost function:
\begin{equation}
    \Psi(Y) = \sum_i |\vec{Y}_i - \sum_j W_{ij}\vec{Y}_{j}|^2.
\end{equation}
During the optimization the $\vec{Y}_i$ coordinates are adjusted, and the $W_{ij}$ weights are kept fixed.

We used both normal, and modified versions of LLE, which are implemented in \texttt{scikit-learn} \citep{scikit}. The modified LLE making use of multiple local weight vectors for each
point, which makes the algorithm much stable \citep{modifiedLLE}. To compute the embedding vectors we only have to manually set the number of neighbors to be considered for each point and a regularization constant, which multiplies the trace of the local covariance matrix of the distances.



\section{Results}
\label{sec:results}

Following \citet{KeplerLLE}, we equally binned the \textit{polyfit} approximations of each light curve into 1000 points before we applied the LLE transformation to reduce the high dimension of the data set. This number of bins is adequate to sample each characteristic feature in case of both contact and detached type phase curves.

We preformed an exhaustive search to find the best manifold technique to keep the similar morphology types as close to each other as possible. Finally, similar to \citet{KeplerLLE}, we found that applying the LLE two times, first, to project the \textit{polyfit} models into a 3D parameter space, then to map the result into a 2D subspace yields the best transformation. Nonetheless, we tried to exchange one of the LLE steps with several other different unsupervised algorithms, such as Isometric Mapping, Spectral Embedding, or simple Singular Value Decomposition (SVD). We found that if not all the morphology types are represented, other methods may perform better. For instance, including SVD in the transformation of W~UMa type light curves resulted in a more informative 2D manifold, where the two axes are \report{correlated} with the characteristics of the primary and secondary minima, respectively. However, when all the types are used, the two steps LLE manifold yields the best low-dimensional parameter space with the smoothest transition between the different subtypes.

 \begin{figure}
   \centering
   \includegraphics[width=\columnwidth]{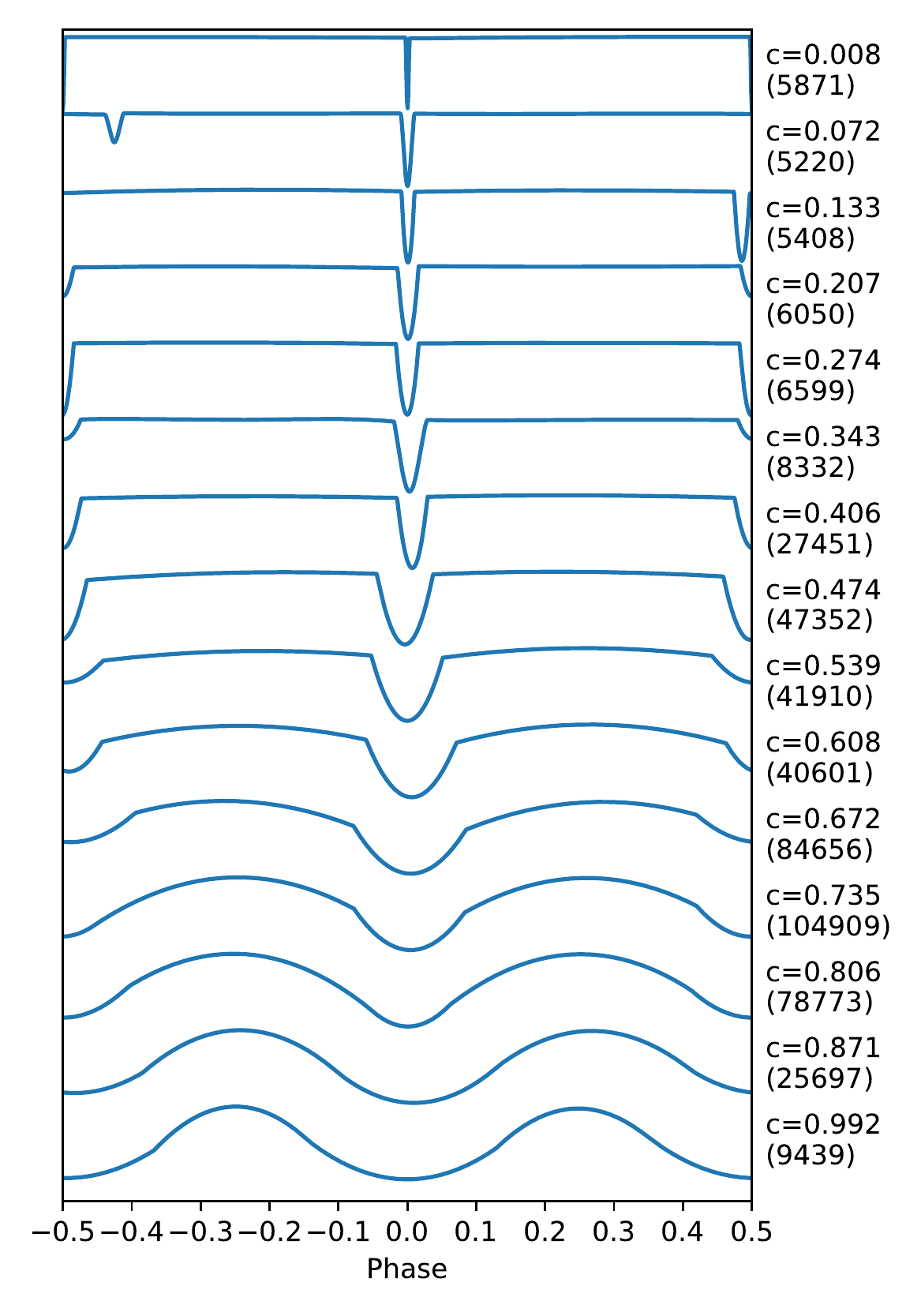}
      \caption{Example \textit{polyfit} model fits as function of the morphology parameter $c$. The $c$ value increases from top to bottom. The numbers in parenthesis show the number of OGLE binary stars with similar light curve shapes. The morphology parameter scales with the ``detachness" of the light curves, i.e. the sum of relative radii of the two components.}
         \label{fig:lc_morphology}
   \end{figure}

Due to the very large size of the OGLE database, we were not able to train the LLE on all \textit{polyfit} models, instead we randomly selected a set of 8735 stars, which is about 1/50 of whole sample, to run the training. To make sure the algorithm is trained on a huge variety of light curve shapes, we ordered the \textit{polyfits} by the periods before the selection, as we assumed that there is a slight correlation between the light curve shape and the orbital period. Fortunately, the LLE can be used to classify previously unseen data after the weights have been adjusted. This property permits the transformation of arbitrary number of \textit{polyfits} without retraining the transformer.

The LLE algorithm has two major parameters that have to be adjusted. We experimentally tested a given range of values and compared the results visually. For both LLE steps, i.e. first mapping from the input data space to 3D, then mapping it to 2D, the number of nearest neighbours $k$, used to reconstruct a given point, was varied between 5--25. The regularization parameter $r$ was tested in the range of $5\cdot10^{-3}-1.5\cdot10^{-2}$. We chose $k=20$ and $k=15$ for the consecutive LLE steps, while both $r$ values was finally set to $1\cdot10^{-2}$. We also experimented with the normal and modified versions of LLE; for first manifold step the former, whilst for the second step the latter has been selected.

The final 3 and 2 dimensional representation of the 1000-bin-long \textit{polyfit} model light curves are plotted in Fig.~\ref{fig:3DLLE}, where we removed outlier points, mainly stars without clearly observed eclipses, where the \textit{polyfit} resulted in a peculiar curve. The upper panel shows three projections of the 3D embedding parameter space. The actual values have no physical meaning alone as they represent the similarities between the neighboring points. We also depicted the type of each object as listed in the OGLE-IV catalog. In the same figure, the lower panel shows the 2D LLE embedding mapped from the previous one. Unfortunately, the detached and semi-detached stars have the same labels, but if we take a look at the light curves \report{corresponding} to the different region of the figure, we will see that the right hand side of the LLE curve is dominated by the Algol type binaries. On the left hand side, the location of those stars that show continuous brightness variation are overlapped. By inspecting the light curves, it can be revealed that the leftmost the point, the more sinusoidal the shape of the curve. There are some stars in the right side of the LLE curve that obviously should not be there. The light curve of these objects are  a combination of a continuous variation and a sharp eclipse and the dominate feature defines the location of the star in the LLE manifold.

As it was recognized in case of the Kepler binaries \citep{KeplerLLE}, the 2D LLE representation can be approximated by a one dimensional curve and the location of each data point along this curve can also be quantified. To do so, we scaled the values along the 1D arc into the range of 0-1 in the following way. First, we measured an angle with respect to an arbitrary point, then using a seventh order polynomial, we adjusted the scale. Then we assigned a single value to each data point based on their distance to the curve by selecting the nearest one. This value is called the morphology parameter $c$ as it scales with the ``detachness" of the eclipses. The ``detachness" is a measure of the sum of the relative radii of the components, thus it can be used as an indicator, however, it should not be represented as an accurate physical unit, because the physics of eclipsing binaries is way more complicated than to be merged into a single parameter. The variation of parameter $c$ along the 2D LLE curve is shown by the labels in the lower panel of Fig.~\ref{fig:3DLLE}. Some example \textit{polyfits} representing the measured light curves that correspond to the different morphology parameters can be seen in Fig.~\ref{fig:lc_morphology}.

The morphology parameter $c$  has been estimated for all the binary stars observed by the primary Kepler mission \citep{KeplerLLE}. We tried to scale our $c$ values as similar as possible to those by determining the morphology of a set of Kepler light curves and comparing them to those listed in the Villanova Kepler Eclipsing Binary catalog \citep{KEBC}, however, our parameters are probably slightly off of those due to two main reasons. Firstly, the LLE algorithm is sensitive to both the initial set of weights and the diversity of the training set. Running the algorithm two times even with the same data set can be resulted in slightly different LLE shapes, especially in the direction of the axes.
Secondly, \citet{KeplerLLE} did not publish the way how they exactly transformed their LLE parameters into a single value.

The presented morphology parameters can be used as a guideline to select different type of eclipsing binaries. From Fig.~\ref{fig:lc_morphology}, it is not easy to put strict limits between the classes, but we can see that the range of $c\leq0.5$ is dominated by detached binaries, the semi-detached systems lie between $0.5<c\leq0.7$, and above $c\sim0.7$ the light curves become more sinusoidal, where the ellipsoidal and overcontact systems can be found. However, utilizing the orbital periods presented in Fig.~\ref{fig:P-c-separate} (and discussed in the next section) might improve the selection process.

 \begin{figure}
   \centering
   \includegraphics[width=\columnwidth]{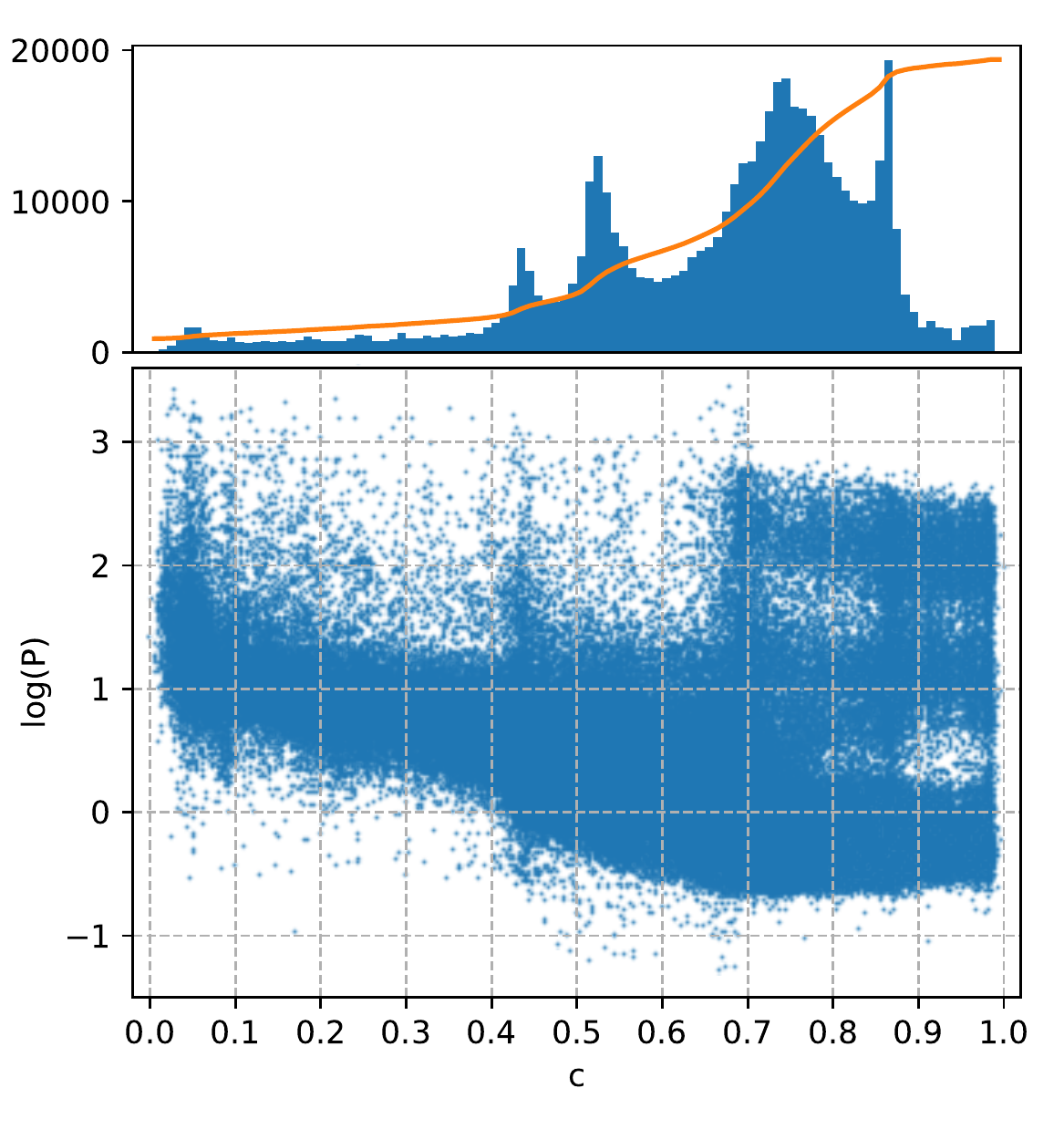}
      \caption{Correlation between the morphology parameter $c$ and the orbital period $P$ (bottom). The top panel shows the distribution of morphology parameter (blue), along with its cumulative histogram (orange).}
         \label{fig:P-c}
   \end{figure}

 \begin{figure}
   \centering
   \includegraphics[width=\columnwidth]{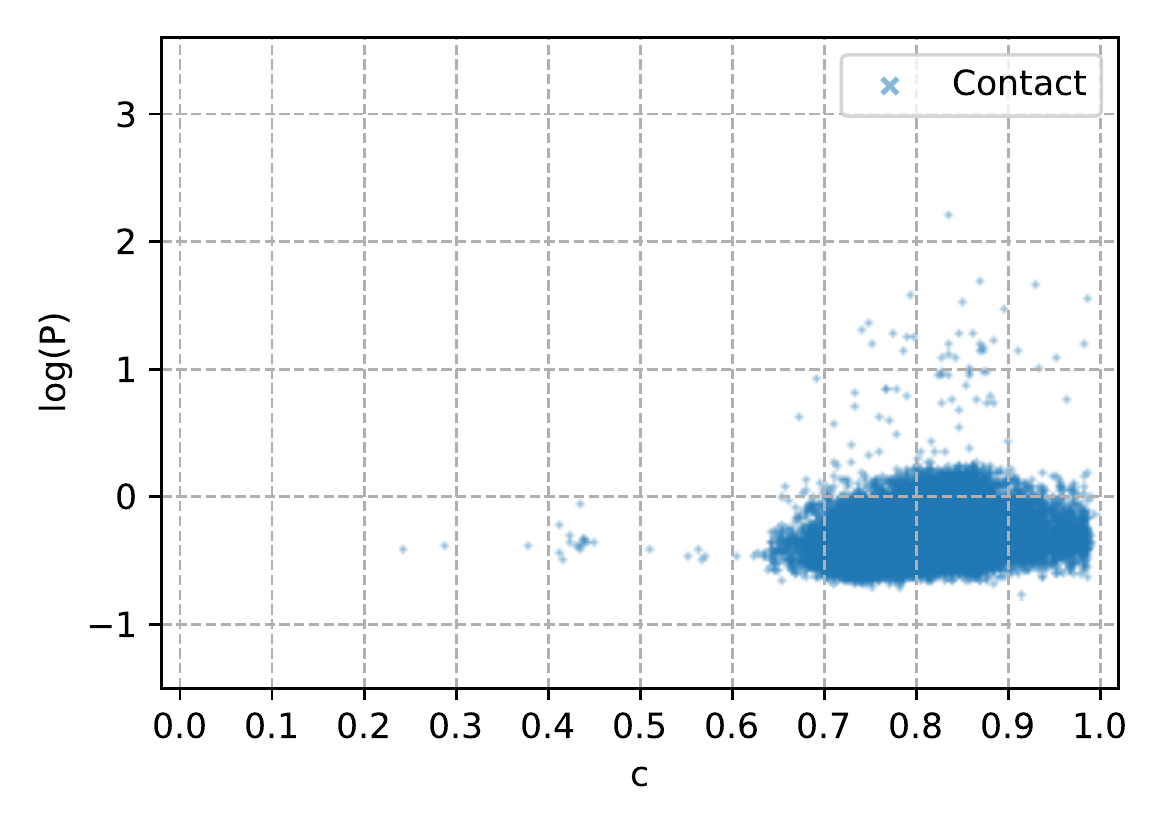}
   \includegraphics[width=\columnwidth]{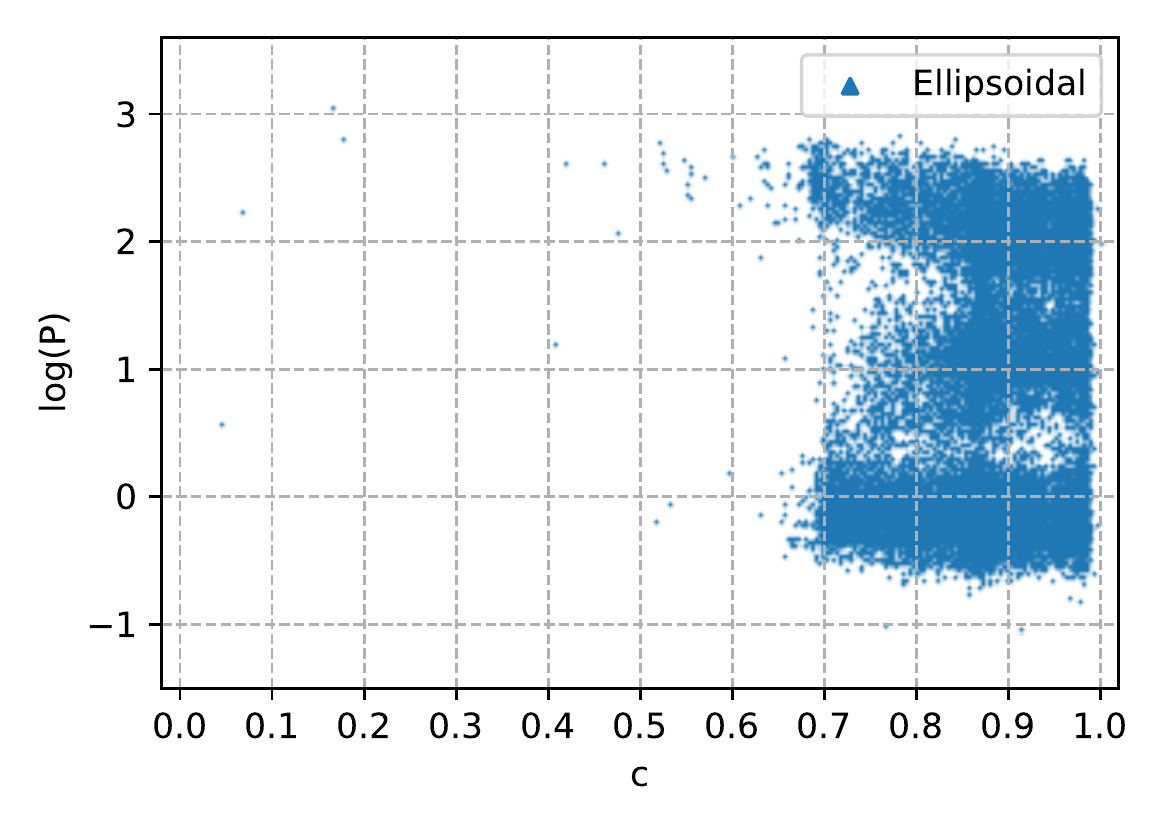}
   \includegraphics[width=\columnwidth]{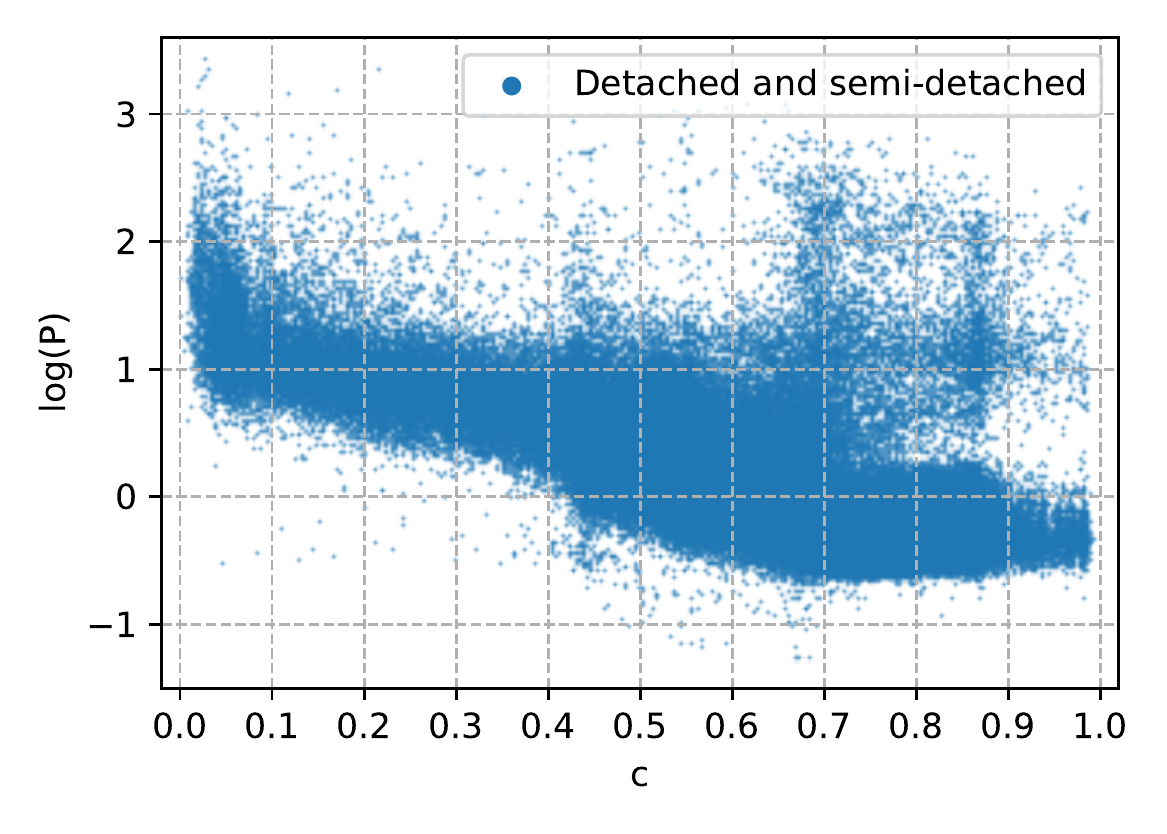}

      \caption{The same correlation between the morphology parameter $c$ and the orbital period $P$ as in Fig.~\ref{fig:P-c}, but for each binary class separately for the galactic bulge sample. From top to bottom, the panels show the location of contact, ellipsoidal, and the detached and semi-detached variables, respectively.}
         \label{fig:P-c-separate}
   \end{figure}

\subsection{Comparison to OGLE classification}

As a correlation is expected between the orbital period $P$ and the morphology parameter $c$, we depicted these parameters against each other in Fig.~\ref{fig:P-c}. The distribution of the bulk data shows several distinct and merged structures. \citet{OGLEBLG} presented the distribution of orbital periods of all the binary stars in the OGLE catalog (see their Fig. 3.). The distribution shows two local maxima, a very sharp one at 0.4~days, which is formed by mainly close main sequence binaries and another one between 100-200~days. The latter probably consists of wide Algol and close red giant systems, such as ellipsoidal variables. In the bottom panel of our Fig~\ref{fig:P-c}, these highlighted features can be clearly recognized, but with more details.

The top panel of Fig.~\ref{fig:P-c} shows the distribution of the morphology parameter along with the cumulative histogram of those values. From this figure, it can be seen that the number of stars per bin is not uniformly distributed, as it is nearly the case for the eclipsing binaries observed by the Kepler telescope (see Figure 4 of \citealt{KeplerLLE}), in contrast, the distribution shows some remarkable peaks. The OGLE binary stars are clumped around some favoured morphology values. However, it should be noted that the direct comparison between the two samples is not possible, as the Kepler \report{K2} data is biased towards short period systems due to the length of those observations.

To study the characteristics of each subtype separately, in Fig.~\ref{fig:P-c-separate}. we shows the correlation between the morphology parameter and the orbital period, from top to bottom, for the contact, ellipsoidal and the detached and semi-detached systems, respectively, in the field of galactic bulge.
Contact stars are clustered in a well defined area with periods shorter than $\sim$1 day and parameter $c$ larger than $\sim$0.65. In contrast, ellipsoidal variables comprise three distinct horizontal strips with average periods of $\gtrapprox$200, $\sim$10 and $\lessapprox$1 days. The first group consists of genuine ellipsoidal variables. The second groups was discovered by \citet{OGLEELL}, who classified those objects based on their location on the Hertzsprung-\report{Russell} diagram as Red Clump stars. As ellipsoidal variables that form the sequence E (introduced by \citealt{SeqE}) in the period-luminosity relation should not have such short periods, the last group probably has been misclassified based on their light curve shapes that show alternating minima. Unfortunately, detached and semi-detached binary systems are not distinguished in the OGLE-IV catalog, thus they can be found everywhere along the whole parameter space. Nonetheless, seemingly these stars form two merged features. A horizontal strip with period shorter than $\sim$1~day and $c$ value greater than $\sim$0.6. Presumably, these stars are semi-detached variables with continuous brightness variation and alternating minima. The other feature lies mainly between periods $\sim$1-100~days with two tails at the edges. This should be formed by Algols, where less detached systems tend to have shorter periods. The well separated Algols, with probably elliptical orbits, have the longest orbital periods, much longer than 10~days. Finally, those points that lie outside of these regions correspond to stars with light curves dominated by reflection effect or misclassified objects.

 \begin{figure}
   \centering
   \includegraphics[width=\columnwidth]{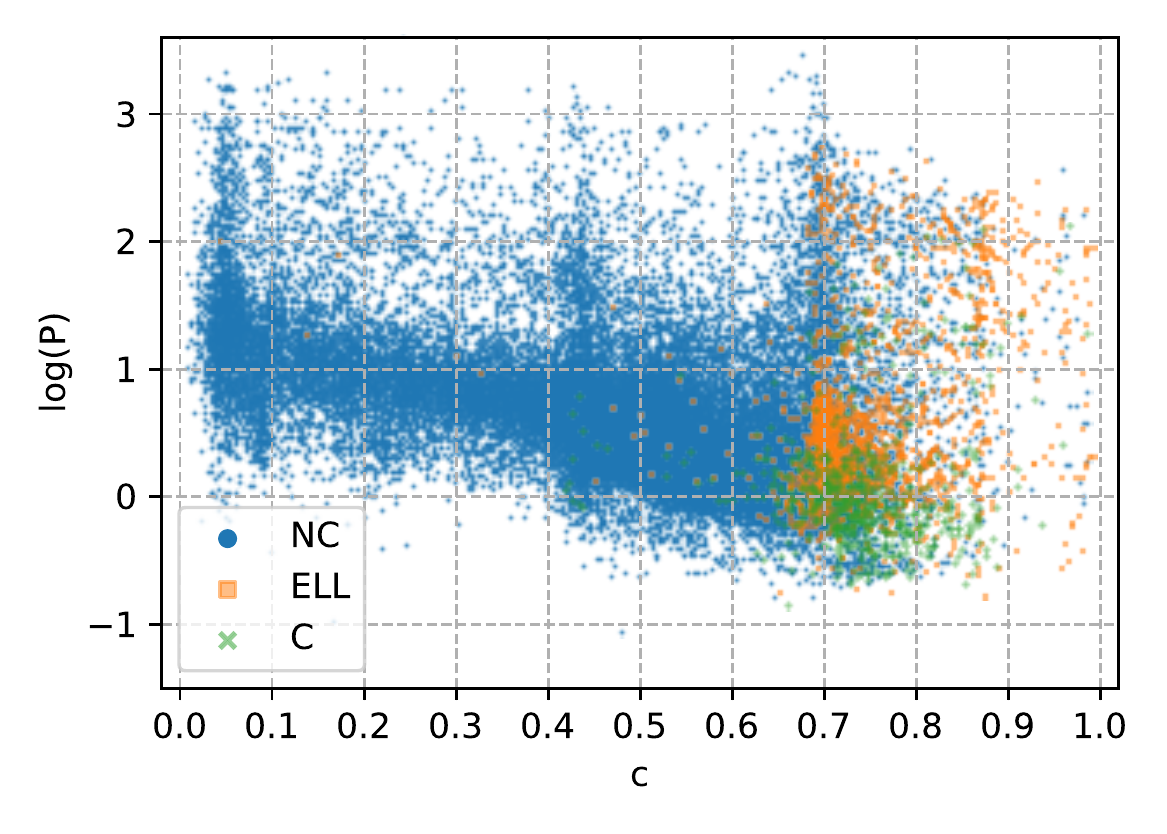}
      \caption{The same correlation between the morphology parameter $c$ and the orbital period $P$ as in Fig.~\ref{fig:P-c}, but only for the Magellanic Clouds sample. The blue points, orange squares and green crosses depict the detached and semi-detached (NC), ellipsoidal (ELL), and contact (C) variables, respectively.}
         \label{fig:P-c_lmcsmc}
   \end{figure}

The morphology parameter -- orbital period correlation for the binary stars in the field of Magellanic Clouds (MCs) is shown in Fig.~\ref{fig:P-c_lmcsmc}. The most notable difference compared to the sample of galactic bulge stars is the very low number of objects that show sinusoidal-like variability, that is, have large morphological values. In the short orbital period range, these are mainly binaries consists of W UMa type stars, which are missing from the OGLE catalog, due their low brightness, which are typically fainter than 21 mag, and the catalog only contains stars brighter than 20.4 mag in I-band \citep{OGLEMC}. In the long period regime stars are usually red giants that show ellipsoidal variation and have intrinsically large luminosities. The low number of this kind of objects can probably be explained by the age of the population. As it was revealed by \citet{OGLEMC} from the position of stars in the color-magnitude diagram, the majority of MC binaries are main sequence stars belonging to the young population, thus the lack of low mass evolved giants. As it can be seen, this kind of behaviour can be revealed with the help of our automatic morphological classification.

\section{Data and Code Availability}

Table~\ref{tab:morph}. lists all the newly derived parameters, i.e. the fine-tuned periods and times of minimum brightness, and the estimated morphology classifications ($c$ values) for each star. The full version of this table is available in a machine-readable format in the online journal.

Moreover, we decided to turn our code into a user-friendly pipeline, which can be used not only to reproduce our results, but to apply the modified version of \textit{polyfit} and the morphological classification to other large set of binary star light curves. \report{The software is available on GitHub\footnote{\url{https://github.com/astrobatty/polyfitter}.} under an MIT License and version 1.0 is archived in Zenodo \citep{polyfitter}.}

\begin{table*}
    \centering
    \caption{List of 449\,706, 40\,163 and 8\,399 OGLE eclipsing binaries in the field of galactic bulge, Large-, and Small Magellanic Clouds, respectively, comprising the final morphology classifications, along with the newly derived times of minimum brightness (T$_0$) and fine-tuned orbital periods (P). The full version of this table is available in a machine-readable format in the online journal, with a portion shown here for guidance regarding its form and content.
    }
    \label{tab:morph}
    \begin{tabular}{c c c c}
\hline\hline
ID & T$_{0,\rm{new}}$ & P$_{\rm{new}}$ & Morphology \\
 & (HJD-2450000) & (days)  &  \\ \hline
 OGLE-BLG-ECL-000001 & 7000.03429 & 0.2090391 & 0.658\\
 OGLE-BLG-ECL-000002 & 7000.0068 & 0.213671 & 0.76\\
 OGLE-BLG-ECL-000003 & 7000.1715 & 0.2143425 & 0.752\\
 OGLE-BLG-ECL-000004 & 7000.1587 & 0.2148819 & 0.683\\
 OGLE-BLG-ECL-000005 & 7000.2059 & 0.2154964 & 0.674\\
 $\cdots$  & $\cdots$  & $\cdots$  & $\cdots$ \\
\hline
    \end{tabular}
\end{table*}

\section{Summary}
\label{sec:summary}

In this work we trained an unsupervised classification algorithm, Local Linear Embedding (LLE), to classify each eclipsing binary (EB) star observed by the OGLE survey in the field of galactic bulge and the Magellanic Clouds based on the morphology of light curves. LLE is capable of mapping the high dimensional data space into a low dimensional manifold, while keeping the local geometry and the similarity of adjacent points. We phase folded and fitted the light curves by polynomial chains (\textit{polyfits}), then applied LLE in two consecutive steps to get a 2D representation of 1000-point-long \textit{polyfits}. Finally, we assigned a single value to each star, which correlates well with the ''detachness", i.e. the sum of the relative radii of the components. The morphology parameter, along with the orbital periods, can be used to separate stars belonging to main morphological classes.

We presented the distribution of the morphological values as a function of orbital periods, and compared our results to the Kepler sample of \citet{KeplerLLE}, and showed that the correlation between the two parameters is still clearly detectable in this very large sample of binary systems. We also found a significant difference between the two samples, the distribution of morphology parameters in the OGLE catalog shows noteworthy peaks instead of following a nearly uniform distribution, however, it should be taken into account that that stars in the Kepler EB catalog are biased towards shorter periods due to the limited duration of the observations, while thanks to the 10-year-long duration of the OGLE-IV survey, we were able to extend the findings to the region of much longer orbital periods.

Our code is publicly available to the astronomical community to apply a similar automatic classification to other large catalogs of binary stars.

\acknowledgements
This project has been supported by the Lend\"ulet Program  of the Hungarian Academy of Sciences, project No. LP2018-7/2020.
On behalf of \textit{"Analysis of space-borne photometric data"} project we thank for the usage of MTA Cloud (\url{https://cloud.mta.hu}) that significantly helped us achieving the results published in this paper.

\software{Python \citep{Python3},
Numpy \citep{numpy},
Pandas \citep{pandas},
Scikit-learn \citep{scikit},
Matplotlib \citep{matplotlib}
}

\bibliography{references}{}
\bibliographystyle{aasjournal}



\end{document}